# Optimization of High-Order Quarter-Wave Plate for Residual Birefringence Suppression in FOCS

**Yuechen Liu, Boqi Meng**

***Abstract:*** Fiber optic current sensors (FOCS) are widely adopted in modern power grids due to high sensitivity, excellent insulation, and strong immunity to electromagnetic interference. This prominence necessitates precise investigation into their error sources and corresponding optimization. This study examines reflective FOCS based on the Faraday effect. A theoretical model is established to simulate phase error caused by linear birefringence from the quarter-wave plate. Conventional methods using circular birefringence are analyzed, revealing inherent limitations. Innovatively, a compensation strategy employing high-order quarter-wave plates is proposed to effectively eliminate linear birefringence effects. This approach significantly enhances the accuracy and practicality of FOCS in precision metrology.

## 1. Introduction

Over the past century, global electricity demand has grown exponentially due to industrialization, digitalization, and the integration of renewable energy, highlighting the critical need for stable and secure power system operations to prevent blackouts, economic losses, and safety risks. This context has accelerated the demand for advanced monitoring and protection systems. With ongoing optimization of energy structures and the development of smart grids, higher accuracy, stability, and safety in current measurement technologies are required [1-5]. Current transformers (CTs), essential for current detection, protection, and metering, play a key role in grid safety and fault management. However, increasing grid voltage and current levels expose limitations of conventional CTs, such as magnetic saturation, susceptibility to interference, large size, and high cost, restricting their use in modern smart grids [6-11]. Among emerging alternatives, fiber optic current sensors (FOCS) based on the Faraday effect have gained significant attention for their immunity to electromagnetic interference, compact design, and suitability for high-voltage environments [12-14]. Faraday-based FOCS primarily include interferometric Sagnac loops and in-line structures, with the latter being more widely studied. Progress has been made in modulation/demodulation techniques and interference suppression, including dual-sensor configurations for simultaneous current and vibration measurement, and reflective structures that improve resistance to vibration and temperature variations [15-20]. Artificial neural networks have also been applied to reduce system noise [19-23], though training complexity limits their practical adoption. Despite these advances, a major remaining challenge remains inherent vulnerability and limited robustness of the quarter-wave plate (QWP), a critical optical component.

In the in-line FOCS structure, a quarter-wave plate (QWP) is commonly employed to convert linearly polarized light into circularly polarized light to mitigate the effects of linear birefringence. However, the fabrication and assembly processes of the waveplate inevitably introduce additional linear birefringence, which leads to measurement errors in the system output and degrades measurement accuracy. This has become one of the primary sources of error in FOCS.

In this work, we presented a comprehensive theoretical and numerical analysis aimed at mitigating phase errors in FOCS induced by linear birefringence originating from QWP. Conventional methods, such as employing spun fibers to introduce substantial circular birefringence for compensation, suffer from structural complexity and poor stability under practical operating conditions, particularly in dynamic environments affected by temperature fluctuations and mechanical stresses. To overcome these constraints, we propose an innovative compensation strategy that incorporates high-order quarter-wave plates, specifically engineered to suppress linear birefringence more effectively than traditional solutions. Consequently, this investigation is dedicated to the integration of a high-order QWP within an in-line FOCS architecture to address this persistent challenge. Compared with traditional solutions, the high-order QWP, designed with specific spinning functions (e.g., linear or cosine gradient profiles) and optimized spinning rates, can more effectively counteract phase errors caused by linear birefringence, thereby significantly improving measurement accuracy and operational stability. By bridging theoretical insights with practical design innovations, this work provides a foundational framework for future developments in high-performance fiber-optic sensing technologies.

## 2. In-line FOCS and Error caused by QWP

The optical configuration of the in-line FOCS is illustrated in Figure 1. The system integrates several critical optoelectronic components, including a broadband light source, a series of polarization control and modulation elements, a sensing fiber coil, a reflective termination, and an optical signal detection unit. The operational sequence begins with the emission of unpolarized broadband light from a superluminescent diode or similar high-coherence source. This unpolarized radiation is first transmitted through an initial linear polarizer, which selectively transmits light oriented along a specific polarization axis, thereby producing a well-defined linearly polarized beam. The polarized light subsequently traverses a 45° fusion splice—an intentionally misaligned joint that introduces a controlled angular offset between the principal axes of the connecting fiber segments. This deliberate misalignment serves to evenly distribute optical power between the two orthogonal polarization modes of the fiber.

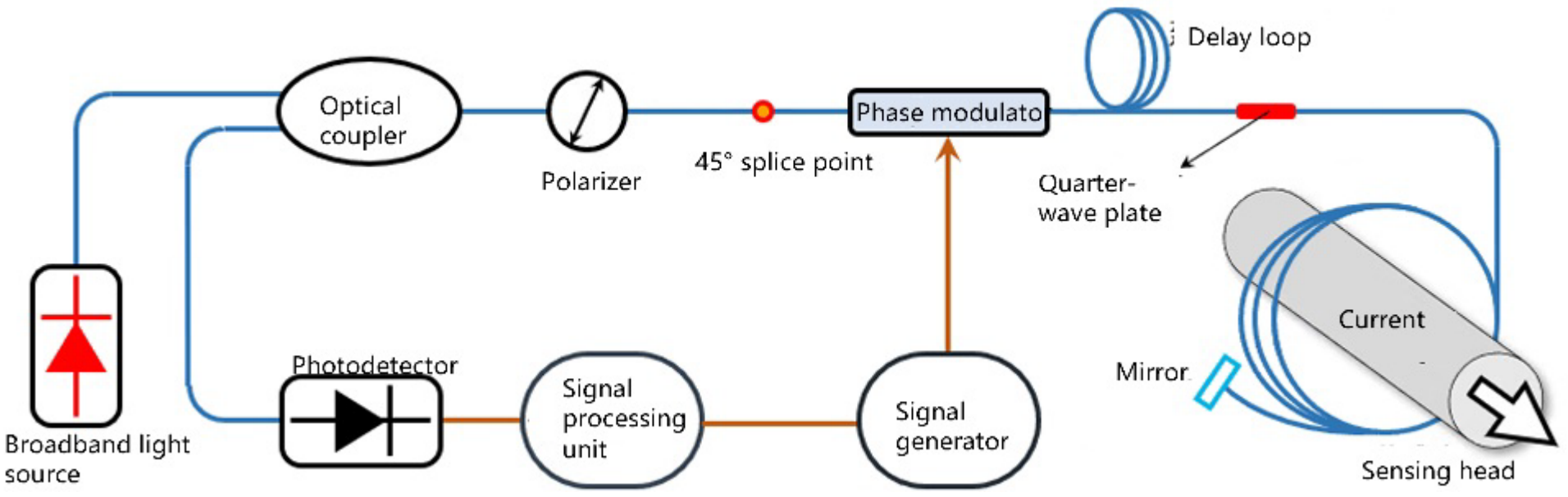

Figure 1 Setup of the FOCS

Following this, the light enters a QWP, a crucial polarization conditioning element that introduces a phase shift of approximately π/2 between the fast and slow axes. Through this retarding action, the QWP effectively transforms the incident linearly polarized light into two counter-rotating circularly polarized waves, specifically left-hand and right-hand circular polarizations (LHCP and RHCP). The Jones matrix of each components can be expressed as follows [3,4]:

Polarizer:

$$L_p = \begin{bmatrix} 1 & 0 \\ 0 & 0 \end{bmatrix} \tag{2.1}$$

45° fusion point

$$L_{45°\mathrm{in}} = \frac{\sqrt{2}}{2}\begin{bmatrix} 1 & 1 \\ -1 & 1 \end{bmatrix} \tag{2.2}$$

QWP:

$$L_{\lambda/4in} = \frac{\sqrt{2}}{2}\begin{bmatrix} 1 & i \\ i & 1 \end{bmatrix} \tag{2.3}$$

$$L_{\lambda/4out} = \frac{\sqrt{2}}{2}\begin{bmatrix} 1 & -i \\ -i & 1 \end{bmatrix} \tag{2.4}$$

Sensing ring:

$$L_{Fin} = \begin{bmatrix} cosF & -sinF \\ sinF & cosF \end{bmatrix} \tag{2.5}$$

$$L_{Fout} = \begin{bmatrix} cosF & sinF \\ -sinF & cosF \end{bmatrix} \tag{2.6}$$

Fiber mirror:

$$L_{mirror} = \begin{bmatrix} 1 & 0 \\ 0 & 1 \end{bmatrix} \tag{2.7}$$

Therefore, the final electric field going through all the fiber components is:

$$E = L_p L_{45°\text{out}} L_{\lambda/4out} L_{Fout} L_{mirror} L_{Fin} L_{\lambda/4in} L_{45°\text{in}} L_p E_{in} \tag{2.8}$$

And the output of the FOCS should be:

$$I_{out} = E^2[1 + \cos(4F)] \tag{2.9}$$

In this scheme, the FOCS will output error results if the QWP deviates from ideal design. The Jones matrix of the waveplate must be modified as:

$$L_{\frac{\lambda}{4}} = \cos\frac{\rho}{2} \begin{bmatrix} 1 + i\tan\frac{\rho}{2}\cos 2\beta & i\tan\frac{\rho}{2}\sin 2\beta \\ i\tan\frac{\rho}{2}\sin 2\beta & 1 - i\tan\frac{\rho}{2}\cos 2\beta \end{bmatrix} \tag{2.10}$$

In this equation, $\rho$ represents the phase retardation introduced by the waveplate, and $\beta$ represents the splicing angle between the elliptical PMF and the panda-type PMF. Moreover, $\rho$ is equal to $\Delta nd$ with $\Delta n$ representing the birefringence in PMF and $d$ is the cutting-length. In order to visualize the error, we substitute the Jones matrix of the practical quarter-wave plate into the previous derivation and compare the relative error($error = \frac{I_{out}-I_{ideal}}{I_{ideal}} \cdot 100\%$) between the practical light intensity and the ideal light intensity.

When this imperfect wave plate acts on the input optical field:

$$\begin{bmatrix} 1 \\ 0 \end{bmatrix}$$

the output Jones vector is given by:

$$E_{out} = \begin{bmatrix} \cos\frac{\rho}{2} + i\,sin\frac{\rho}{2}cos\,2\beta \\ i\sin\frac{\rho}{2}\sin 2\beta \end{bmatrix} \tag{2.11}$$

The actual output intensity can thus be expressed as:

$$I_{out} = {E_{out}}^2[1 + \cos(4F)] \tag{2.12}$$

Subsequently, the relative error is defined as:

$$error = \frac{I_{out} - I_{ideal}}{I_{ideal}} \cdot 100\% \tag{2.13}$$

which serves as an important metric for evaluating device imperfections. For typical parameters,

namely a phase deviation of ±500 μm and an angle deviation of ±2°, the interference contrast is found to decrease by approximately 3%. This result corresponds to the error curve shown in Fig. 2. It is therefore evident that even minor deviations in cutting length and splicing angle can lead to significant performance degradation, underscoring the necessity of strictly controlling fabrication precision.

For this analysis, we only consider the following deviations:

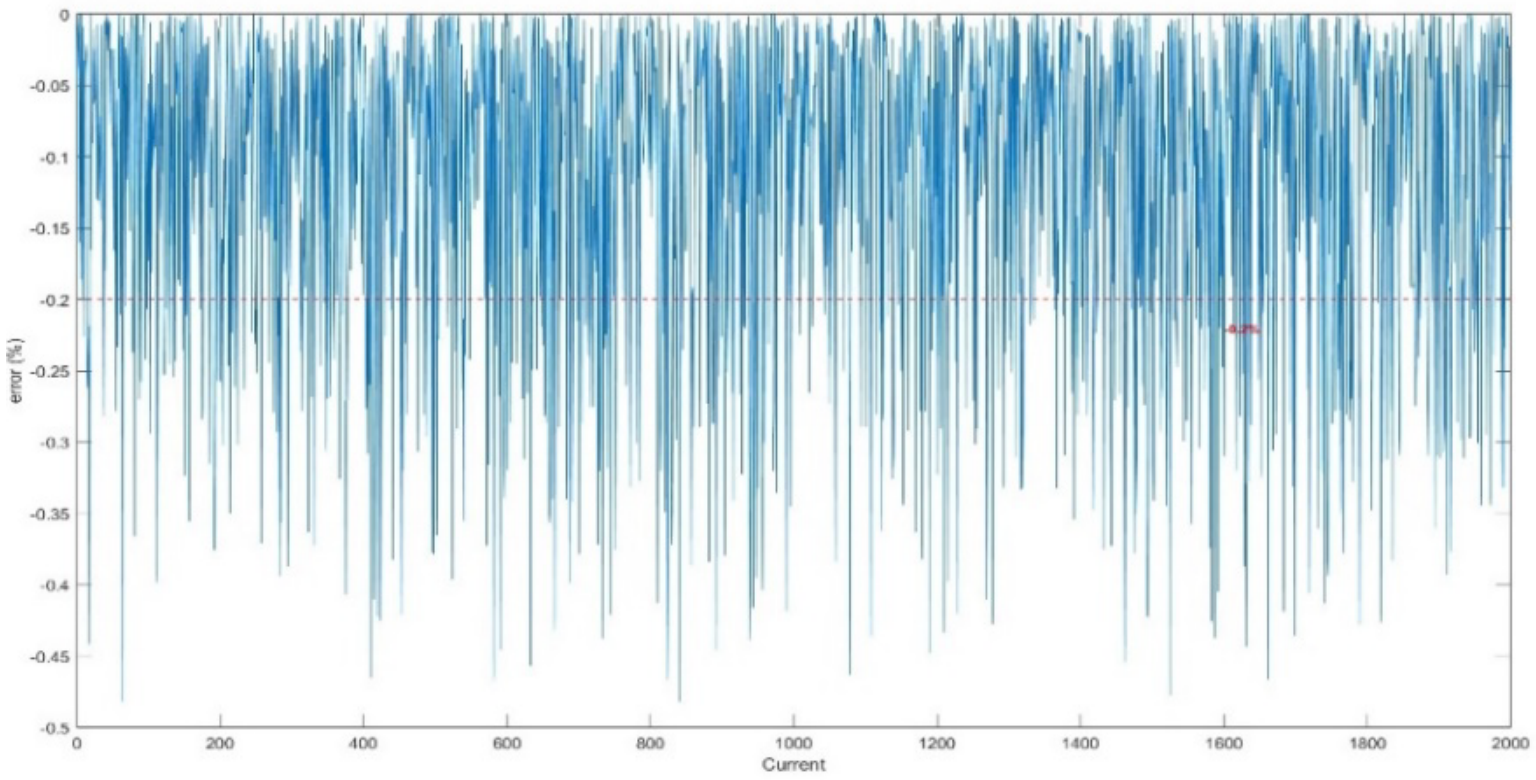

Figure 2 FOCS output error caused by manufacturing imperfections

The error is shown in Figure 2. For typical parameters, namely a phase deviation of ±500 μm and an angle deviation of ±2°, the interference contrast is found to decrease by approximately -3%. This result corresponds to the error curve shown in Figure 2.

In engineering practice, the error of a standard sensor is specified to not exceed 0.2%. As can be seen from the diagram, the error introduced by the quarter-wave plate is significantly larger than this limit. Thus, eliminating the linear birefringence caused by the misalignment of the quarter-wave plate has become an urgent issue. To mitigate the QWP error, some publications use spun fiber [7-8] which, however, may render the FOCS system more complex. We simulated and compared the output light intensities under ideal, twisted, and non-twisted fiber conditions using MATLAB, as illustrated in Figure 3.

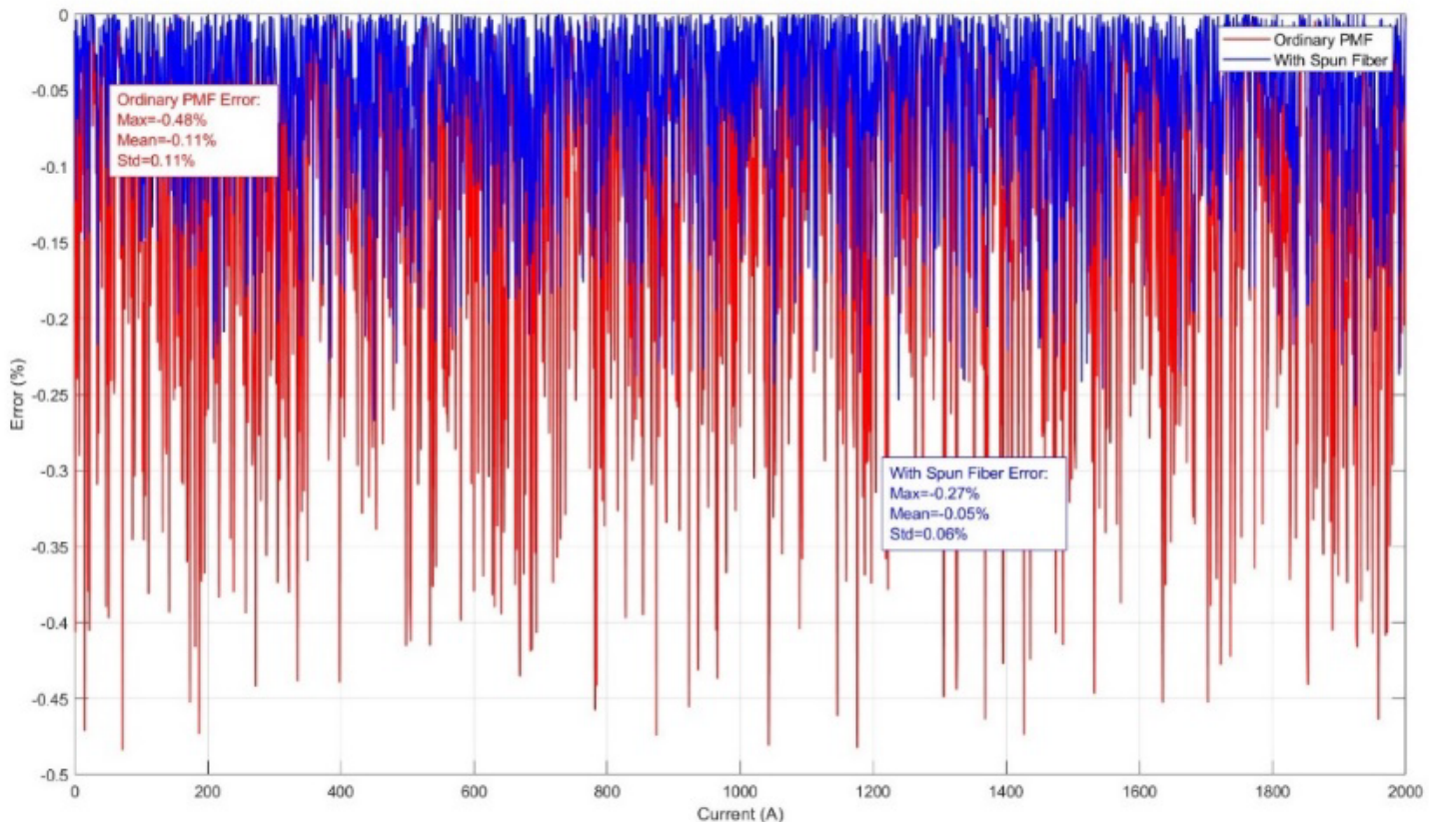

Figure 3 Comparison of FOCS errors when using PMF or spun fiber

## 3. High-order QWP

### 3.1 Basic concept

To mitigate the above-mentioned deficiencies, scientists proposed an all-in-one sensing head based on a high-order QWP [12-16] (which is also call spun quarter wave plate, SQWP). The polarization-maintaining fiber (PMF) is subjected to high temperature and rotated from its slow axis to fast axis orientation until reaching the maximum spinning rate $\xi_{max}$ after which it is rotated uniformly. Specifically, the rotation mechanism employs a servo or stepper motor with a high-resolution encoder for precision control, increasing rotational speed under a predefined function (e.g., linear or cosine) with gradual acceleration to minimize stress. Adjustment steps are kept within ±0.1 rpm and response time within 10–50 ms, while closed-loop feedback (e.g., PID) limits angular deviation to ±0.01°. Temperature stability is maintained at 20–25 °C via thermostatic or water-cooling systems, monitored with ±0.5 °C accuracy. Symmetric heating and preheating mitigate thermal non-uniformity, and stretching halts beyond ±2 °C deviation. Conventional fiber-based quarter-wave plates (QWPs) rely on beat length ($L_b$, typically mm–cm in PMFs) for phase retardation, requiring mm–cm lengths and extreme fabrication precision to avoid phase errors. In contrast, high-order QWPs (SQWP or PPT[24]) operate via pulsed polarization transformation, with effective lengths in the decimeter range (10–30 cm or more). They avoid strict $L_b$ matching, maintain standard fiber diameter (~125 μm) for compatibility, and tolerate greater length variation—enabling easier fabrication, enhanced installation flexibility, and improved engineering feasibility in optical communication systems.

While coupled mode theory (CMT)—solvable numerically via the Runge-Kutta method—describes electric field evolution within local coordinates, a more intuitive visualization of polarization evolution in global coordinates is needed for the high-order QWP. Microscopically, spun structures like spun fiber (SF) and high-order QWP can be modeled as a superposition of numerous infinitesimal retarders. This concept has been validated through the differential element method framework.

Considering light propagation along a polarization-maintaining fiber, the polarization state of a high-order quantum wavefront can be described by the Jones vector $E(z)$. The propagation relation at an arbitrary position $z$ along the fiber is given by:

$$E(z + \Delta z) = J(z, \Delta z)E(z) \tag{3.1}$$

where $J(z, \Delta z)$ is the Jones matrix of a fiber segment of length $\Delta z$. Dividing the total fiber length

$L$ into $N$ small segments, each of length $\Delta z = \frac{L}{N}$, the overall propagation matrix can be expressed as a finite product:

$$J_{total} = \prod_{k=1}^{N} J(z_k, \Delta z)\,, z_k = (k-1)\Delta z \tag{3.2}$$

As $\Delta z \to 0$ (i.e., $N \to \infty$), this finite product approaches the form of an infinite product:

$$J_{total} = \prod_{n=1}^{\infty} J(z_n, \Delta z) \tag{3.3}$$

which represents the mathematical expression of the "infinite product of Jones matrices."

Assuming that each small-segment matrix can be expanded as $J(z_k, \Delta z) = I + A(z_k)\Delta z + O(\Delta z^2)$, where $A(z)$ characterizes the local polarization properties of the fiber and $I$ is the identity matrix, the error between the finite product and the infinite product mainly arises from the accumulation of higher-order terms $O(\Delta z^2)$:

$$Error \sim N \cdot O\left(\left(\Delta z^2\right) = N \cdot O\left(\frac{L^2}{N^2}\right) = O\left(\frac{L^2}{N}\right) \tag{3.4}$$

Thus, the error is inversely proportional to $N$. To keep the error below a given tolerance $\epsilon$, the segment number should satisfy:

$$N \gtrsim \frac{L^2}{\epsilon} \tag{3.5}$$

In practical fiber simulations and high-order QWP design, numerical calculations indicate that when $N = 1000$, the error is typically less than 0.1%, which meets engineering accuracy requirements. Consequently, the overall Jones matrix of the high-order QWP is expressible as [21-22]:

$$J_{\text{high-order QWP}} = \lim_{N\to\infty} J_N J_{N-1} \cdots J_i \cdots J_2 J_1 \tag{3.6}$$

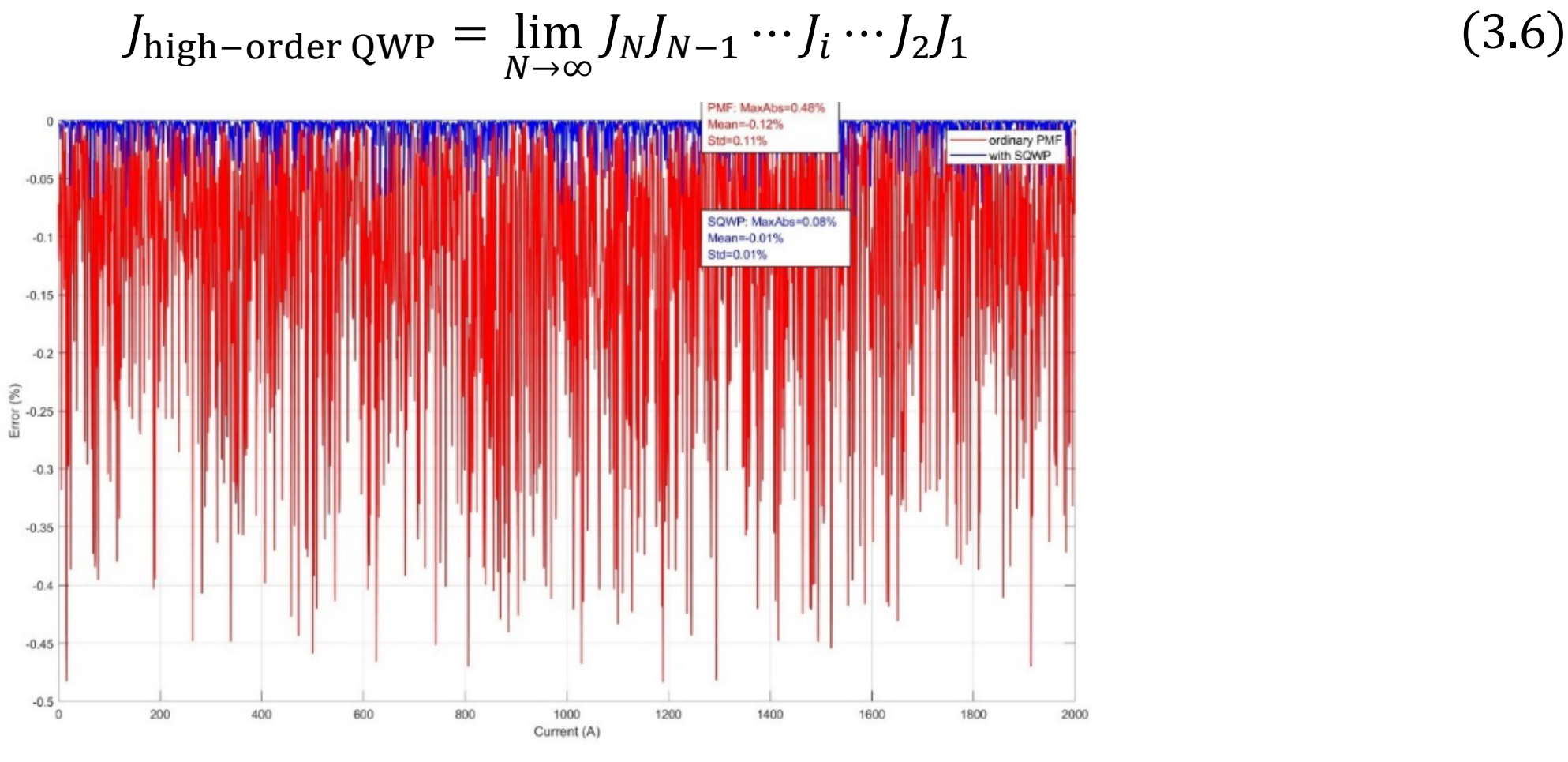

Figure 4 Comparison of FOCS errors when using PMF or high-order QWP (SQWP)

Quantitative results show that within the 0–2000 A range, ordinary PMF exhibits a maximum absolute error of 0.48% and an average error of -0.12%. After introducing high-order QWP, these values decrease to 0.08% and -0.01%, respectively, demonstrating over 80% error reduction and stable improvement across current intervals due to effective suppression of birefringence and polarization angle deviations. Through theoretical, numerical, and experimental comparisons with conventional spun fibers (SFs), high-order QWPs show error variation <0.1%/°C (20–25°C), SOP fluctuation <0.2%, and insertion loss <0.2 dB, outperforming SFs which exhibit higher error variation (0.3–0.5%/°C), SOP fluctuation >0.5%, and insertion loss of 0.3–0.5 dB under varying temperature or prolonged operation. Microscopically, both structures model superpositions of infinitesimal retarders, but high-order QWPs, with smoothly varying cosine rotation rates, enable precise retardation control and stable polarization conversion, whereas SFs suffer from localized stress and rotational gaps causing error accumulation. Overall, high-order QWPs provide superior polarization conversion accuracy, lower loss, better thermal stability, and greater long-term reliability, indicating enhanced engineering adaptability for high-speed optical communications, fiber sensing, and precision polarization control applications.

**3.2 Wave plate optimization**

In the practical fabrication of a high-order QWP, the rotational rate applied during the fiber drawing process is strategically engineered to escalate progressively from zero to a predetermined maximum value. This progression is governed by a specifically selected predefined function—typically linear or cosine in form (Linear:$\xi = \xi_{max} \cdot (z - L_1)/L_2$;Cosine: $\xi = \xi_{max} \cdot (0.5 - 0.5\cos(\pi(z - L_1)/L_2))$ ;Here, $\xi$ represents the rotation rate and z the positional coordinate along the fiber axis.)—which plays a critical role in determining the efficiency and uniformity of optical polarization conversion [10-13]. The choice of rotation function directly influences the alignment of anisotropic axes within the fiber, thereby affecting the cumulative retardance and ultimately the polarization behavior of the resulting waveplate.

A fundamental metric for evaluating the performance of such devices is the **polarization ellipticity**, defined as the ratio of the minor axis to the major axis of the polarization ellipse. The ellipticity can be calculated using the Jones matrix. If the Jones vector of the light is：

$$E = \begin{bmatrix} E_x \\ E_y \end{bmatrix} \tag{3.2}$$

then the ellipticity is expressed as：

$$\varepsilon = \frac{|E_y|}{|E_x|} \tag{3.3}$$

Ellipticity, a key indicator of polarization quality, measures how closely the polarization state approaches circular polarization (value of 1), with higher values denoting greater stability. Significant deviation from 1 suggests instability induced by linear birefringence or rotation rate fluctuations. Fluctuations along the fiber can be quantified by the derivative of the rotation rate function, dξ/dz, where higher values indicate abrupt variations leading to mode coupling and polarization-dependent loss, while lower values correspond to smoother transitions and improved performance. Thus, precise design and control of the rotation function—supported by real-time monitoring and feedback—are essential to minimize ellipticity variations and meet performance requirements in applications such as optical sensing and telecommunications. As shown in Figure 5, the left and right sides correspond to linear and cosine rotation functions, respectively. The upper plots illustrate the 3D evolution of the electric field vector along the fiber length (0 to –300 mm), showing the gradual transformation of the polarization state from linear to elliptical/circular.

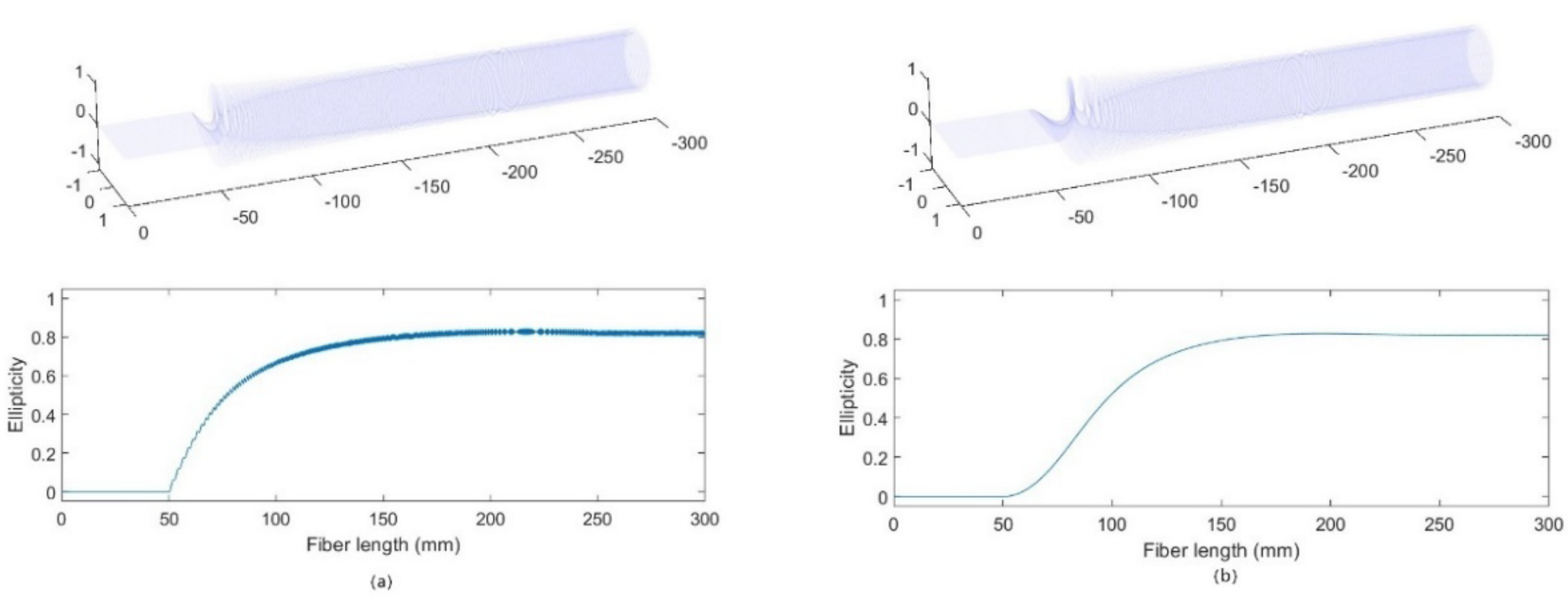

Figure 5 Polarization evolution within the high order wave plate

In the simulation, the parameter $\frac{\xi}{\delta}$was employed to represent different levels of $\xi_{max}$, where $\xi_{max} = \frac{\xi}{\delta} * \delta$) .It should be emphasized that the parameter $\delta$ denotes the intrinsic birefringence-related phase retardation per unit length, which originates from the inherent material anisotropy of the fiber. Thus, normalizing the rotation amplitude by $\delta$ allows different cases of $\xi_{max}$ to be compared under a unified physical scale. The choice of $\frac{\xi}{\delta} = 1,3,5,10$ corresponds to gradually increasing rotation strengths relative to the intrinsic birefringence, thereby enabling a clear assessment of the trade-off between conversion efficiency and stability.

In the simulation, four representative values of $\frac{\xi}{\delta} = 1,3,5,10$ were chosen to systematically evaluate the impact of rotation modulation amplitude on the stability of the fiber ellipticity. The rationale is as follows:

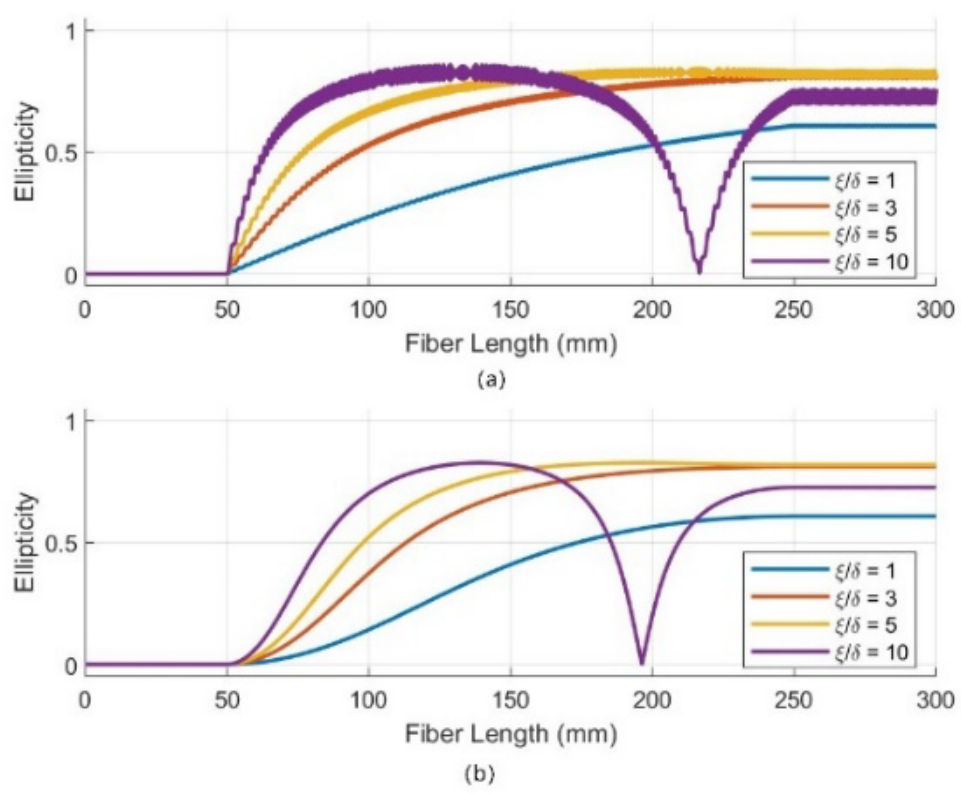

Figure 6 Relationship between different spinning functions and different values of $\frac{\xi}{\delta}$

First, when $\frac{\xi}{\delta} = 1$ , the modulation amplitude is of the same order as the intrinsic birefringence of the fiber. This represents a weak modulation or baseline reference case, providing a direct comparison with the material's inherent properties. Second, for $\frac{\xi}{\delta} = 3$ and $\frac{\xi}{\delta} = 5$ the modulation amplitude is significantly larger than the intrinsic birefringence, corresponding to medium-to-strong modulation regimes. These cases reflect practical operating conditions where the trade-off between conversion efficiency and stability becomes apparent. Finally, when $\frac{\xi}{\delta} = 10$, the modulation amplitude is much stronger than the intrinsic birefringence, representing an extreme modulation case. This allows us to explore the stability limits of the system under extreme conditions and to provide boundary references for engineering design.In summary, the gradual increase of $\frac{\xi}{\delta}$ from weak to extreme modulation ensures comprehensive coverage of different operating regimes, avoids biased analysis, and facilitates a thorough discussion of the balance between conversion performance and stability.

Furthermore, the maximum rotation speed, $\xi_{max}$, significantly influences system behavior. Since the Jones matrix of a high-order QWP can be expressed as an infinite product of incremental wave plate matrices, a higher $\xi_{max}$ accelerates the conversion from linear to circular polarization, enhancing operational efficiency.

However, this comes at the expense of increased ellipticity fluctuation, as $\Delta\varepsilon \propto \left|\frac{d\xi}{dz}\right|$. Let the ideal ellipticity trajectory be $\varepsilon_0(z)$ and the actual trajectory be $\varepsilon(z)$, then:

$$\Delta\varepsilon(z) = \varepsilon(z) - \varepsilon_0(z) \tag{3.4}$$

Perturbation theory indicates that this deviation is proportional to the nonuniformity of the rotation rate, i.e., $\Delta\varepsilon \propto \left|\frac{d\xi}{dz}\right|$. On the other hand, the overall magnitude of the rotation speed $\xi_{max}$ also amplifies this deviation, since under the same derivative condition, a larger rotation amplitude implies a faster acceleration of the coordinate system, leading to stronger instability. Therefore:

$$\Delta\varepsilon \propto \left|\frac{d\xi}{dz}\right| \xi_{max} \tag{3.5}$$

Numerical simulations conducted in MATLAB, employing the normalized parameter ξ/δ to represent different $\xi_{max}$ levels, corroborate this trade-off.

For the linear rotation function, ellipticity rises rapidly over the initial 50–250 mm segment, with higher $\frac{\xi}{\delta}$ values yielding steeper gradients—a consequence of the constant derivative driving swift polarization evolution. Beyond the peak, a slight decline or fluctuation occurs due to phase accumulation overshoot, particularly pronounced at elevated $\frac{\xi}{\delta}$ values, reflecting inherent instability from abrupt transitions. The cosine function, by comparison, produces a smooth sigmoidal trajectory in ellipticity. Although higher $\frac{\xi}{\delta}$ steepens the curve, the sinusoidal derivative ensures milder fluctuations. Nonetheless, a post-peak decline emerges at high $\frac{\xi}{\delta}$ levels, resulting from phase mismatch and accelerated polarization transition, which disrupts the birefringence balance.

From a practical standpoint, elevated $\xi_{max}$ values facilitate rapid state of polarization (SOP) conversion, beneficial in high-speed optical communications and sensor responsiveness. However, this advantage is counterbalanced by increased manufacturing challenges and susceptibility to environmental perturbations. To verify the theoretical derivation and determine the optimal range of $\xi_{max}$ , we fabricated and tested actual high-order QWPs. Comparative experiments were conducted for $\frac{\xi}{\delta} = 2,3,4,5,6$, with evaluation metrics including peak-to-peak ellipticity fluctuation $\Delta\varepsilon_{pp}$, root-mean-square error (RMS), and fabrication difficulty (e.g., fiber draw uniformity and rotational control precision). The results are as follows:

- For $\frac{\xi}{\delta} = 2$, ellipticity fluctuations were minimal, but the conversion efficiency was relatively low;
- For $\frac{\xi}{\delta} = 3 - 5$, ellipticity fluctuations remained well controlled, while conversion efficiency was significantly improved, and fabrication difficulty remained within acceptable limits;
- For $\frac{\xi}{\delta} = 6$ , although the conversion speed was faster, ellipticity fluctuations increased substantially ($\Delta\varepsilon_{pp} > 1\%$), and fabrication difficulty and process stability were significantly compromised.

Thus, a balanced approach advocating $\xi_{max} \approx 3–5\delta$, combined with a cosine rotation pattern, is recommended to harmonize efficiency and stability. Future refinements may involve the synthesis of higher-order continuous functions to enhance derivative smoothness and further suppress ellipticity fluctuations. It is imperative to note that these findings derive from idealized simulations; practical implementations must account for additional variables such as wavelength drift and thermal variations, which may exacerbate ellipticity perturbations.

## 4. Discussion and conclusion

### 4.1 Discussion

This investigation delineates several inherent limitations that merit careful consideration. A primary constraint lies in the intricate fabrication process of high-order QWP, which imposes exceptionally stringent demands on rotational rate control, cutting precision, and splice quality. To quantitatively evaluate the influence of these environmental factors on the QWP, simulations were conducted considering typical wavelength drift (±10 nm) and temperature fluctuation (±20 °C). The performance metrics were defined as the peak-to-peak ellipticity fluctuation:

$$\Delta\varepsilon_{pp} = max_{\varepsilon(z)} - min_{\varepsilon(z)} \tag{3.6}$$

and the root-mean-square (RMS) ellipticity variation:

$$RMS_{\varepsilon} = \sqrt{\frac{1}{L}\int_{0}^{L}(\varepsilon(z) - \bar{\varepsilon})^{2}dz} \tag{3.7}$$

Simulation results indicate that wavelength drift can increase $\Delta\varepsilon_{pp}$ by approximately 5–8% and $RMS_{\varepsilon}$ by about 3%, while temperature fluctuations can increase $\Delta\varepsilon_{pp}$ by approximately 10–12% and $RMS_{\varepsilon}$ by about 5–7%. These omissions highlight a significant gap between theoretical models and operational reality.

To address these challenges, future research should prioritize several strategic directions: (1) the development of more advanced fabrication techniques for high-order QWP, including the exploration of higher-order continuous rotation functions to optimize polarization stability and reduce insertion losses; (2) an examination of hybrid configurations that integrate high-order QWP with twisted fibers or other polarization-manipulating elements, accompanied by rigorous empirical validation to quantify performance under varied environmental conditions; and (3) the incorporation of computational approaches such as neural networks or adaptive control algorithms to facilitate real-time compensation of environmental noise, thereby augmenting measurement accuracy and operational robustness. Implementing these initiatives would significantly advance the practical deployment of high-order QWP in sensitive optical systems.

### 4.2 Conclusion

This study addresses the mitigation of linear birefringence errors in FOCS, focusing on the limitations of conventional quarter-wave plates. Using Jones calculus and a theoretical model, we demonstrate that manufacturing and environmental factors introduce linear birefringence exceeding the 0.2% error tolerance. While twisted fibers provide partial compensation through circular birefringence, practical issues such as splice losses and stability limit their use in high-precision applications. As a superior alternative, high-order QWP are proposed. These achieve uniform birefringence via controlled rotational drawing, effectively converting linear to circular polarization and suppressing linear birefringence. Numerical analyses confirm that high-order QWP reduce errors and avoid fusion-splice-related inaccuracies. A comparison of linear and cosine rotation functions shows that linear profiles offer uniformity despite end-point discontinuities, whereas cosine functions enable smoother polarization conversion. Optimizing the maximum rotation rate $\xi_{max}$ remains crucial for balancing conversion efficiency and ellipticity stability. In summary, high-order QWP provides a robust solution for linear birefringence suppression in FOCS, with significant implications for smart grid applications. Future work on process optimization and composite design promises further performance enhancements, supporting reliable smart grid monitoring and safety.